\documentstyle[12pt]{article}
\def\case#1/#2{\textstyle\frac{#1}{#2} }
\begin{document}
\title{Integrability Conditions for Irrotational Dust with a Purely
Electric Weyl Tensor: A Tetrad Analysis}
\author{ {\sc W. M. Lesame$^{1,2}$, P. K. S. Dunsby$^{1}$
and G. F. R. Ellis$^{1}$}\\
\normalsize{$1$ \it Department of Applied Mathematics,
University of Cape Town, }\\
\normalsize{ \it Rondebosch 7700, Cape Town, South Africa.} \\
\normalsize{$2$ \it Department of Applied Mathematics, University of Fort
Hare,} \\
\normalsize{\it Private Bag X1314, Alice 5700, South Africa.}}
\date{$\mbox{}$ \vspace*{0.3truecm} \\ \normalsize  \today}
\maketitle
\thispagestyle{empty}
\begin{abstract}
All spacetimes for an irrotational collisionless fluid with a purely electric
Weyl tensor, with spacetime curvature determined by the exact Einstein
field equations, are shown to be integrable. These solutions include the
relativistic generalisations of the Zeldovich solutions of Newtonian theory.
Thus our result shows the consistency of various studies of "silent"
universes (where such consistency was assumed rather than proved).
 \end{abstract}
\vspace*{0.2truecm}
\begin{center}{\it Subject headings:}\\
cosmology\,-\,galaxies: clustering,
formation\,-\,hydrodynamics\,-\,relativity\,-\,exact solutions
\end{center}
\newpage
\section{Introduction}
The dynamics of self\,-\,gravitating collisionless matter (dust)
that is irrotational and with vanishing magnetic part of the Weyl tensor
has been the focus of interest in the last couple of years
\cite{bi:MPS1,bi:MPS2,bi:BMP}. \\

The assumption  that the matter
can be described by a collisionless fluid is reasonable since
we wish to study cosmological perturbations during the matter dominated era.
However, because of the high degree of non\,-\,linearity
in Einstein's field equations, a number of additional physically motivated
approximations have been introduced to make the problem more tractable. \\

The second assumption is that the fluid flow is irrotational. Physically this
is reasonably well justified provided the scales of interest are not too small
(i.e. where rotation starts to become important). Then the kinematical
condition that $\omega_{ab}=0$ implies the flow is
hypersurface orthogonal. Furthermore if it is initially irrotational then it
will remain irrotational in the absence of dissipative effects.
It is the third assumption of vanishing ``magnetic'' part
of the Weyl tensor that has led to most discussion.\\

A remarkable feature of perfect fluid spacetimes with vanishing vorticity
$\omega_{ab}$ and ``magnetic'' part of the Weyl tensor $H_{ab}$ is that they
admit an orthonormal tetrad associated with the matter  4\,-\, velocity $u^a$
which is simultaneously an eigen\,-\,tetrad for the shear of the matter flow
$\sigma_{ab}$ and the ``electric'' part of the Weyl tensor $E_{ab}$
\cite{bi:BR}. It follows that, apart from two special cases, coordinates exist
in which the metric $g_{ab}$ and the tensors $\sigma_{ab}$ and $E_{ab}$
are all diagonal. This together with the further condition that the flow is
geodesic $\dot{u}_a=0$ leads to a considerable simplification of the
propagation equations, reducing them to a set of six ordinary differential
equations (the Raychaudhuri and continuity equations  together with two pairs
of equations for the independent components of $\sigma_{ab}$ and $E_{ab}$)
\cite{bi:MPS1}. In addition to these propagation equations various constraint
equations have to be satisfied that result from these conditions. Apart from
the equations that directly express the restriction $H_{ab}=0$, there
is an additional constraint which follows from the $\dot{H}$
Bianchi identity. Physically this can be thought of as limiting the spatial
variation of the ``electric'' tidal field $E_{ab}$.  \\

When these constraint equations are satisfied, each fluid
element evolves independently of each other until the formation of caustics,
when the one\,-\,to\,-\,one mapping between fluid elements and space points is
lost.  Such a universe model was dubbed  {\it silent} by Matarrese {\it et.
al.} \cite{bi:MPS2}, because no information can be exchanged between the
fluid elements which was not already present in the initial conditions
(which must be chosen to satisfy the constraint equations).\\

The aim of this paper is to show that the required integrability conditions
are satisfied: the time-propagation equations are consistent with the
constraints. Two of the constraints are clearly sensitive to the assumption of
$H_{ij}=0$ and warrant a full consistency check: {\it viz} i) the usual
$H=0=w$ constraint which restricts the spatial distribution of the shear
$\sigma_{ij}$; ii) and the additional constraint due to $\dot{H_{ij}} =0$
described above. In fact in showing that in the case of dust, vectors of
the tetrad are hypersurface orthogonal, Barnes and Rowlingson (1989)
\cite{bi:BR} performed
a consistency test on the part of the $\dot{H_{ij}}$ constraint for which
$i=j$. We carry out a full consistency on all the constraints.\\

Sections \ref{sec:prop} and \ref{sec:cons} list the evolution and the
constraints equations in covariant form. Results of calculation of the time
derivative of constraints are given in section \ref{sec:timeprop}, using a
tetrad formalism. A sample calculation appears in the appendix. Consistency
analysis is performed first for the trivial Type O fields, which evolve
as Friedmann-Robertson-Walker (FRW) models. The type D and type I fields are
tested separately. In both cases, integrability conditions are consistently
satisfied.\\

Kinematic and tetrad notation used here are the same as in Ellis (1967)
\cite{bi:ellis1}. Latin indices run from $0$ to $3$ and Greek indices from $1$
to $3$. The tetrad $\{e_{0},e_{i}\}$ is chosen with the timelike
vector $e_{0}$ chosen as the fluid flow vector, and the associated parameter
$\tau$ measuring proper time between the surfaces orthogonal to the fluid
flow (spanned by the vectors $e_{i}\}$). This can be done since for dust we
have zero acceleration.  Such a specialization simplifies time propagation
calculations.

\section{Relativistic dynamics of irrotational dust} \label{sec:dyn}
In this section we will give a brief summary of the relativistic dynamics
of irrotational dust [3] in terms of covariant variables that represent the
observable kinematical and dynamical quantities [4], focusing on the case of
irrotational dust with vanishing ``magnetic part'' of the Weyl
tensor: $p=\omega_{ab}=H_{ab}=0$.

\subsection{Variables}
The relativistic dynamics of dust is determined by the Einstein Field
equations and by the continuity equation for the matter\-\,
stress--energy--tensor $T_{ab}=\rho u_au_b$, where $\rho$ is the
energy density and
$u^a$ is the normalized 4\,-\,velocity of the fluid ($u^au_a=-1$). At each
spacetime point we can define a projection tensor $h_{ab}=g_{ab}+u_au_b$
for which $h_{ab}u^a=0$. With $u^a$ and $h_{ab}$ it is possible to split
the covariant derivative of any tensorial quantity into a time derivative
and a spatial derivative. In particular the first covariant derivative
of $u^a$ can be written as $u_{a;b}=v_{ab}+a_au_b$, where
$v_{ab}\equiv h^c{}_ah^d{}_bu_{c;d}$ is the spatial part satisfying
$v_{ab}u^b=0$, and $a^a\equiv\dot{u}^a\equiv u^a{}_{;b}u^b$ is the
acceleration vector which is orthogonal to the fluid flow ($a^au_a=0$).
It is standard to split $v_{ab}$ into three kinematical quantities:
it's trace $\Theta\equiv v^a{}_{;a}$, symmetric, trace\,-\,free part
$\sigma_{ab}\equiv v_{(ab)}-\case{1}/{3}h_{ab}\Theta$, and antisymmetric
part $\omega_{ab}\equiv v_{[ab]}$. The physical meaning of these quantities
are as follows: $\Theta$ measures the volume expansion or contraction
of the fluid, $\sigma_{ab}$ is the shear tensor which describes the rate
at which a spherical fluid volume is distorted into an ellipse and
$\omega_{ab}$ is the vorticity tensor giving the rate of rotation with
respect to a local inertial frame.

\subsection{Kinematic equations}
For a perfect fluid the fluid acceleration is only determined by pressure
gradients so the restriction of vanishing pressure implies that
$a^a=0$. This means that each fluid element moves along a geodesic.
The conservation of energy and momentum $T^{ab}{}_{;b}=0$
leads to only one further equation, the continuity equation:
\begin{equation}
\dot{\rho}=-\rho\Theta\;.
\end{equation}
With the second restriction of vanishing vorticity $\omega_{ab}=0$,
the equations for the kinematic quantities follow from the Ricci identity:
$u_{a;d;c}-u_{a;c;d}=R_{abcd}u^b$ [4,5]. The expansion scalar $\Theta$ obeys
the Raychaudhuri equation:
\begin{equation}
\dot{\Theta}+\case{1}/{3}\Theta^2+2\sigma^2+\case{1}/{2}\kappa\rho=0\;,
\end{equation}
where $\sigma^2\equiv\case{1}/{2}\sigma^{ab}\sigma_{ab}$ is the shear
scalar and $\kappa=8\pi G$ is the gravitational constant. Note that in the
homogeneous and isotropic case where $\sigma_{ab}=0$ and $\Theta=3H$,
this equation reduces to the familiar Friedmann equation:
\begin{equation}
3\dot{H}+3H^2+\case{1}/{2}\kappa\rho=0\;.
\end{equation}
The only other kinematic equation is for the shear. It is given by:
\begin{equation}
\dot{\sigma}_{ab}+\sigma_{ac}\sigma^c{}_b-\case{2}/{3}\sigma^2h_{ab}
+\case{2}/{3}\Theta\sigma_{ab}+E_{ab}=0\;,
\end{equation}
where $E_{ac}=E_{(ac)}\equiv C_{abcd}u^bu^d$ is the ``electric'' part of
the Weyl tensor $C_{abcd}$ (satisfying $E_{ac}u^c=0$, $E^a{}_a = 0$).
$E_{ab}$ is that part of the gravitational field which describes tidal
interactions. The Weyl tensor can be decomposed into $E_{ab}$ and another
tensor called the ``magnetic'' part: $H_{ac} =
H_{(ac)} \equiv \case{1}/{2} \eta_{ab}{}^{gh} C_{ghcd}u^bu^d$ (satisfying
$H_{ac}u^c = 0$, $H^a{}_a=0$). This is the part of the gravitational field
that describes magneto-gravitic effects, and allows gravitational waves.\\

So far we have explicitly made two assumptions, those of vanishing pressure
and vorticity. In what follows we will make the third and most important
assumption, namely, neglecting the influence of the ``magnetic'' part of the
Weyl tensor. This implies neglecting the interaction of gravitational
waves with the system.

\subsection{Constraint equations}
Besides the evolution equations for the kinematical equations,
there are several constraints that our variables must satisfy.
On setting $p=\omega_{ab}=H_{ab}=0$ we obtain as non-trivial constraints,
\begin{equation}
h^e{}_b\left(\case{2}/{3}\Theta^{;b}-h^d{}_c\sigma^{bc}{}_{;d}\right)=0\;,
\end{equation}
the `$(0,\nu)$' field equations, and
\begin{equation}
h^t{}_ah^s{}_d\sigma_{(t}{}^{b;c}\eta_{s)fbc}u^f=0\;,
\end{equation}
the condition that $H_{ab} = 0$. Additionally there is a Friedmann like
equation giving the Ricci-scalar of the 3-spaces orthogonal to $u^a$;
this is a first-integral of (1) and (2).

\subsection{Bianchi Identities}
Additionally, the Bianchi identities must be satisfied, as they are the
integrability conditions for the other equations. With our restrictions,
they take the form:
\begin{equation}
h^t{}_ah^d{}_sE^{as}{}_{;d}=\case{1}/{3}h^t{}_b\rho^{;b}\;,
\end{equation}

\begin{equation}
h^m{}_ah^t{}_c\dot{E}^{ac}+h^{mt}\sigma^{ab}E_{ab}+\Theta E^{mt}
-3E_s{}^{(m}\sigma^{t)s}=-\case{1}/{2}\rho\sigma^{tm}\;,
\end{equation}
the `div $E$' and `$\dot{E}$' equations respectively, and

\begin{equation}
\eta^{tbpq}u_b \sigma^d{}_p E_{qd}= 0\;,
\end{equation}
and
\begin{equation}
h_a{}^{(m}\eta^{t)rsd}u_rE^a{}_{s;d}=0\;.
\end{equation}
which are the `div $H$' and `$\dot{H}$' equations respectively in the
case where $H_{ab}$ vanishes.

\section{Tetrad approach} \label{sec:prop}
A direct conversion of these equations into a tetrad system which is an
eigenframe for both the shear tensor and the  Weyl tensor yields the following
time-evolution equations (Barnes and Rowlingson, 1989 \cite{bi:BR});
\begin{eqnarray}
\dot{\rho} &=& - \rho \theta\;, \label{eq:in1}\nonumber\\
\dot{\theta}&=&- \case{1}/{3}\theta^{2}
- (\sigma_{1}{}^{2} + \sigma_{2}{}^{2} + \sigma_{3}{}^{2})
-\case{1}/{2}\rho\;,
\label{eq:in2} \nonumber\\
\dot{\sigma}_{\mu}&=&-(\sigma_{\mu})^{2} -\case{2}/{3}\theta\sigma_{\mu}
+\case{1}/{3}(\sigma_{1}{}^{2} + \sigma_{2}{}^{2} + \sigma_{3}{}^{2})
- E_{\mu}\;, \label{eq:in3} \nonumber\\
\dot{E}_{\mu}&=&-\theta E_{\mu} -\case{1}/{2} \rho \sigma_{\mu}
+ 3\sigma_{\mu}E_{\mu} - (\sigma_{1}E_{1} + \sigma_{2}E_{2} +
\sigma_{3}E_{3})\;, \label{eq:in4}
\end{eqnarray}
being respectively the tetrad forms of (1), (2), and of the diagonal parts
of (4) and (8). The non-diagonal parts of (4) and (8) (the propagation
equations for $\dot{E}_{ab}$ and $\dot{\sigma}_{ab}$ with
$a\neq b$) introduces two  additional conditions;
\begin{eqnarray}
0&=&(\sigma_{\nu}-\sigma_{\mu})\Gamma^{\nu}{}_{0\mu}\;,
\;\;\;\mu\neq \nu\;, \label{eq:in5} \\
0&=&(E_{\nu}-E_{\mu})\Gamma^{\nu}{}_{0\mu}\;\;\;\mu\neq \nu\;,
\label{eq:in6}
\end{eqnarray}
where the Ricci rotation coefficients are defined by
$\Gamma_{abc} := e_{a}. \bigtriangledown_{b} e_{c}$ [3], and one can raise
and lower indices in the first place as usual by use of the metric.

\subsection{Uniqueness of tetrad}
The tetrad used here is an orthonormal tetrad. If we denote the local
coordinate system by $\{x^{i}\}$ and the tetrad by $\{{\bf e}_{a}\}$ then the
equations (Ellis, 1967 \cite{bi:ellis1})
\begin{equation}
{\bf e}_{a} = e_{a}^{i}(\partial/\partial x^{i})
\label{eq:in6a}
\end{equation}
define the functions $e^{i}_{a}$, which are components of the tetrad
vectors ${\bf e}_{a}$ with respect to the basis $\partial/\partial x^{i}$, and
are also directional derivatives of the coordinate functions $x^{i}$ as
\begin{equation}
e^{i}_{a} = \partial_{a}(x^{i}).
\label{eq:in6b}
\end{equation}

The choice of a tetrad which simultaneously diagonalizes the shear
tensor and the
Weyl tensor possess the following uniqueness properties:
\begin{enumerate}
\item The tetrad is uniquely determined for the two cases:
\begin{enumerate}
\item Distinct $E_{\mu}$  and/or distinct $\sigma_{\mu}$ values.
\item  Degenerate shear tensor and  Weyl tensor, with the principal
planes not coinciding i.e., (say) $E_{1}=E_{2}\neq E_{3}$ and $
\sigma_{1}=\sigma_{3}\neq
\sigma_{2}$. Here equations (\ref{eq:in5}) and (\ref{eq:in6}) yield
\begin{equation}
\Gamma^{\nu}{}_{0\mu} = 0; \;\;\;\; \mu \neq \nu\;.
\label{eq:in7}
\end{equation}
\end{enumerate}
\item The tetrad is free by a rotation in (say) the $e_{1},e_{2}$
plane if both the shear and the Weyl tensor are degenerate in the
same plane i.e., $E_{1}=E_{2}\neq E_{3}$ and $\sigma_{1}=\sigma_{2}\neq
\sigma_{3}$.

We write the tetrad freedom as
\begin{eqnarray}
\bar{e}^{i}_{1}&=&e^{i}_{1}\cos \phi +e^{i}_{2} \sin \phi\;,
\label{eq:in8} \nonumber\\
\bar{e}^{i}_{2}&=&-e^{i}_{1}\sin \phi +e^{i}_{2} \cos \phi;,
\;\;\;\;\;\; \phi=\phi(x^{0},x^{i}) \label{eq:in9}
\end{eqnarray}
and from (\ref{eq:in5}) or (\ref{eq:in6}) we have
\begin{equation}
\Gamma^{3}{}_{02} =  \Gamma^{3}{}_{01} = 0\;.
\label{eq:in10}
\end{equation}
A rotation of the tetrad can then be performed to determine $\partial\phi
/\partial x^{0}$ from the requirement that
\begin{equation}
\Gamma^{2}{}_{01} = 0\;.
\label{eq:in11}
\end{equation}
This leaves the tetrad arbitrary by an initial rotation in a surface
$x^{0}= const$. The  remaining freedom is later used to
set $\Gamma^{1}{}_{32}=0$ in (\ref{eq:sp4}) below.
\end{enumerate}
Equation (\ref{eq:in7}) is thus valid generally and simplify calculations
greatly. Both equations (\ref{eq:in5}) and (\ref{eq:in6}) are satisfied.
Also it follows trivially from the propagation equations  that
$E_{\mu} =0$ if and only if $\sigma_{\mu}=0$ (FRW model).

\section{Constraint Equations} \label{sec:cons}
The constraint equations give spatial restrictions on the tetrad form
of the dynamical variables. The $(0,\nu)$ field equations (5) are:
\begin{eqnarray}
\case{2}/{3}\partial_{1}\theta&=&\partial_{1}\sigma_{1} +
(\sigma_{1}-\sigma_{2})\Gamma^{2}{}_{21} +
(\sigma_{1}-\sigma_{3})\Gamma^{3}{}_{31}\;, \label{eq:con2}\\
\case{2}/{3}\partial_{2}\theta&=&\partial_{2}\sigma_{2}
+ (\sigma_{2}-\sigma_{1})\Gamma^{1}{}_{12}
+ (\sigma_{2}-\sigma_{3})\Gamma^{3}{}_{32}\;, \label{bi:con3}\\
\case{2}/{3}\partial_{3}\theta&=&\partial_{3}\sigma_{3}
+ (\sigma_{3}-\sigma_{1})\Gamma^{1}{}_{13}
+ (\sigma_{3}-\sigma_{2})\Gamma^{2}{}_{23}\;.
\label{eq:con4}
\end{eqnarray}

The `div $E$' equations (7) take the form
\begin{eqnarray}
\case{1}/{3}\partial_{1}\rho&=&\partial_{1}E_{1}+
(E_{1}-E_{2})\Gamma^{2}{}_{21} + (E_{1}-E_{3})\Gamma^{3}{}_{31}\;,
\label{bi:con6} \\
\case{1}/{3}\partial_{2}\rho&=&\partial_{2}E_{2}
+ (E_{2}-E_{1})\Gamma^{1}{}_{12}
+ (E_{2}-E_{3})\Gamma^{3}{}_{32}\;, \label{bi:con7} \\
\case{1}/{3}\partial_{3}\rho&=&\partial_{3}E_{3}
+ (E_{3}-E_{1})\Gamma^{1}{}_{13}
+ (E_{3}-E_{2})\Gamma^{2}{}_{23}\;.
\label{bi:con8}
\end{eqnarray}

The `div $H$' equations (9) are identically satisfied by the
tetrad variables (indeed it is they that allowed us to simultaneously
diagonalise the shear and the Electric part of the Weyl tensor).\\


The equation (6) for vanishing $H_{ab}$ takes the form


\begin{eqnarray}
\Gamma^{1}{}_{32}(\sigma_{2}-\sigma_{1})&=&
\Gamma^{1}{}_{23}(\sigma_{3}-\sigma_{1})\;, \label{eq:con11} \\
\Gamma^{2}{}_{31}(\sigma_{1}-\sigma_{2})&=&
\Gamma^{2}{}_{13}(\sigma_{3}-\sigma_{2})\;, \label{eq:con12} \\
\Gamma^{3}{}_{12}(\sigma_{2}-\sigma_{3})&=&
\Gamma^{3}{}_{21}(\sigma_{1}-\sigma_{3})\;,
\label{eq:con13}
\end{eqnarray}
\begin{eqnarray}
\partial_{1}(\sigma_{3}-\sigma_{2})&=&
 \Gamma^{3}{}_{31}(\sigma_{1}- \sigma_{3})
-\Gamma^{2}{}_{21}(\sigma_{1} -\sigma_{2})\;, \label{eq:con14} \\
\partial_{2}(\sigma_{3}-\sigma_{1})&=&
\Gamma^{3}{}_{32}(\sigma_{2}-\sigma_{3})
-\Gamma^{1}{}_{12}(\sigma_{2}-\sigma_{1})\;, \label{eq:con15} \\
\partial_{3}(\sigma_{2}-\sigma_{1})&=&
\Gamma^{2}{}_{23}(\sigma_{3}-\sigma_{2})
-\Gamma^{1}{}_{13}(\sigma_{3}-\sigma_{1})\;.
\label{eq:con16}
\end{eqnarray}

Finally the dynamical restriction $\dot{H}_{\mu\nu}=0$, equation (10),
introduces constraints given by


\begin{eqnarray}
\Gamma^{1}{}_{32}(E_{2}-E_{1})&=&
\Gamma^{1}{}_{23}(E_{3}-E_{1})\;, \label{eq:zap2} \label{eq:con18} \\
\Gamma^{2}{}_{31}(E_{1}-E_{2})&=&
\Gamma^{2}{}_{13}(E_{3}-E_{2})\;, \label{eq:zap3} \label{eq:con19} \\
\Gamma^{3}{}_{12}(E_{2}-E_{3})&=&
\Gamma^{3}{}_{21}(E_{1}-E_{3})\;,
\label{eq:con20}
\end{eqnarray}
\begin{eqnarray}
\partial_{1}(E_{3}-E_{2})&=&
\Gamma^{3}{}_{31}(E_{1}- E_{3})-\Gamma^{2}{}_{21}(E_{1} -E_{2})\;,
\label{eq:con21} \\
\partial_{2}(E_{3}-E_{1})&=&
\Gamma^{3}{}_{32}(E_{2}- E_{3})-\Gamma^{1}{}_{12}(E_{2} -E_{1})\;,
\label{eq:con22} \\
\partial_{3}(E_{2}-E_{1})&=&
\Gamma^{2}{}_{23}(E_{3}- E_{2})-\Gamma^{1}{}_{13}(E_{3} -E_{1})\;.
\label{eq:con23}
\end{eqnarray}

\section{Time propagation of constraints} \label{sec:timeprop}
For the propagation equations to be integrable with the chosen restrictions
on the kinematic variables ($w_{\mu\nu}=0=H_{\mu\nu}$), the
constraint equations (\ref{eq:con2}\,-\,\ref{eq:con23}) must be
preserved during the time development of the system. We focus here on the
$\dot{H}$ constraint (32-37); a similar analysis is also valid for constraints
(\ref{eq:con2}\,-\,\ref{eq:con16}).\\

The time propagation of the $\dot{H}$ constraint
in covariant form (10) gives
\begin{equation}
0= h^{(i}{}_{n}\eta^{j)klm}\left[(\dot{E}^{n}{}_{l})_{;m}
-E^{n}{}_{l;p}u^{p}{}_{;m}- R^{n}{}_{mpq}E^{q}{}_{l}u^{p}
+ R^{q}{}_{mpl}E^{n}{}_{q}u^{p} \right]\;.
\label{eq:prop1}
\end{equation}
If we convert  the new constraint (\ref{eq:prop1}) to the tetrad
form we obtain the following. For $\mu=\nu$, we find
\begin{eqnarray}
 0&=&\Gamma^{1}{}_{23}(E_{3}- E_{1})(\sigma_{2}- \sigma_{3})\;,
\label{eq:prop2} \\
 0&=&\Gamma^{2}{}_{31}(E_{1}- E_{2})(\sigma_{1}-\sigma_{3})\;,
\label{eq:prop3} \\
 0&=&\Gamma^{3}{}_{12}(E_{2}- E_{3})(\sigma_{1}-\sigma_{2})\;,
\label{eq:prop4}
\end{eqnarray}
which can be written equivalently as:
\begin{eqnarray}
0&=&\Gamma^{1}{}_{32}(E_{2}- E_{1})(\sigma_{2}-\sigma_{3})\;,
\label{eq:prop5} \\
0&=&\Gamma^{2}{}_{13}(E_{3}- E_{2})(\sigma_{1}-\sigma_{3})\;,
\label{eq:prop6} \\
0&=&\Gamma^{3}{}_{21}(E_{1}- E_{3})(\sigma_{1}-\sigma_{2})\;.
\label{eq:prop7}
\end{eqnarray}
For $\mu\neq \nu$, we get
\begin{eqnarray}
0&=&(E_{2}-E_{3})\partial_{1}\sigma_{3}+
(\sigma_{2}-\sigma_{3})\partial_{1}E_{2}+\case{1}/{3}\Gamma^{3}{}_{31}
(\sigma_{1}-\sigma_{3}) (5E_{1}+4E_{3}) \nonumber\\
&-&\case{1}/{3} \Gamma^{2}{}_{21}(E_{1}-E_{2})
(5\sigma_{1}+4\sigma_{2})\;,
\label{eq:prop8}
\end{eqnarray}
\begin{eqnarray}
0&=&(E_{3}-E_{1})\partial_{2}\sigma_{1}
+ (\sigma_{3}-\sigma_{1})\partial_{2} E_{3}+\case{1}/{3} \Gamma^{1}{}_{12}
(\sigma_{2}-\sigma_{1}) (5E_{2}+4E_{1}) \nonumber \\
&-&\case{1}/{3} \Gamma^{3}{}_{32}
(E_{2}-E_{3}) (5\sigma_{2}+4\sigma_{3})\;,
\label{eq:prop9}
\end{eqnarray}
\begin{eqnarray}
0&=&(E_{1}-E_{2})\partial_{3}\sigma_{2}
+ (\sigma_{1}-\sigma_{2})\partial_{3} E_{1}+\case{1}/{3} \Gamma^{2}{}_{23}
(\sigma_{3}-\sigma_{2}) (5E_{3}+4E_{2}) \nonumber\\
&-&\case{1}/{3} \Gamma^{1}{}_{13}
(E_{3}-E_{1}) (5\sigma_{3}+4\sigma_{1})\;.
\label{eq:prop10}
\end{eqnarray}
(It is not obvious how the curvature terms in (38) cancel in
the transition to the tetrad forms; this is shown in Appendix B).

\section{Specializations}
\subsection{Type O}
For this class, $E_\nu = 0$, all constraint equations are trivially
satisfied except for the Friedmann equation which controls the dynamics,
and the evolution is that of the FRW models.
\subsection{Type D}
Without loss of generality we set $E_{1}=E_{2} = E\neq E_{3}$. The following
tetrad properties are valid:
\begin{enumerate}
\item From equations (\ref{eq:con23}) and (\ref{eq:con18},\ref{eq:con19})
we obtain:
\begin{eqnarray}
\Gamma^{1}{}_{13}&=& \Gamma^{2}{}_{23}\,,
 \label{eq:sp1}\\
\Gamma^{1}{}_{23}&=& \Gamma^{2}{}_{13}=0\;.
\label{eq:sp2}
\end{eqnarray}
\item From either (\ref{eq:con11}) or (\ref{eq:con12}) we write:
\begin{equation}
(\sigma_{2}-\sigma_{1}) \Gamma^{1}{}_{32} =0\;,
\label{eq:sp3}
\end{equation}
which has the following two subcases.
\begin{enumerate}
\item For $\sigma_{2}=\sigma_{1}=\sigma \neq \sigma_{3}$ the tetrad is
free by a rotation in the $e_{1},e_{2}$ plane. As pointed out earlier, a
rotation in that plane can be performed so that equation (\ref{eq:in7})
remains
valid. This leaves the tetrad arbitrary by an initial rotation in a surface
$x^{0}=const$. To determine the tetrad completely we perform a further
rotation of $e_{1}\;,e_{2}$ which preserves (\ref{eq:in11}),
where the value of $\partial \phi/ \partial x^{3}$ is determined from the
requirement that $\Gamma^{1}{}_{32} = 0$ in a surface $x^{0} =const$. From
the Jacobi
identities (\ref{eq:a5}\,-\,\ref{eq:a7}) in appendix it follows that
\begin{equation}
\Gamma^{1}{}_{32} = 0
\label{eq:sp4}
\end{equation}
everywhere. The tetrad vectors can hence be chosen to be hypersurface
orthogonal.
This result was also obtained in \cite{bi:BR}. To complete the
consistency analysis we use the above properties as follows:
\begin{enumerate}
\item Constraints (\ref{eq:con14},\ref{eq:con15}) and (\ref{eq:con21},
\ref{eq:con22}) are written respectively as
\begin{equation}
\partial_{1}\sigma = -\sigma \Gamma^{3}{}_{31}; \;\;\;\;
\partial_{2}\sigma = -\sigma \Gamma^{3}{}_{32}\;,
\label{eq:sp5}
\end{equation}
\begin{equation}
\partial_{1}E = -E \Gamma^{3}{}_{31}; \;\;\;\;
\partial_{2}E = -E \Gamma^{3}{}_{32}\;,
\label{eq:sp6}
\end{equation}
with constraints (\ref{eq:con16}) and (\ref{eq:con23})
identically satisfied due to equation (\ref{eq:sp1}).
\item First we recall that the time propagation of constraints
(\ref{eq:con11}\,-\,\ref{eq:con16}) are identically satisfied. Also
the time propagation equations (\ref{eq:prop2}\,-\,\ref{eq:prop4}) and
(\ref{eq:prop10}) are identically satisfied in this class. The identity
$0=0$ from (\ref{eq:prop10}) follows as expected from the $0=0$ in
(\ref{eq:con23}). On the other hand time propagation equations
(\ref{eq:prop8},\ref{eq:prop9}) take the forms;
\begin{eqnarray}
0&=&-2 E\partial_{1}\sigma+ \sigma\partial_{1}E-\sigma E
\Gamma^{3}{}_{31}\;, \label{eq:sp7} \\
0&=&E \partial_{2}\sigma+2 \sigma\partial_{2}E+\sigma E
\Gamma^{3}{}_{32}
\label{eq:sp8}
\end{eqnarray}
which are identically satisfied on using (\ref{eq:sp5}) and (\ref{eq:sp6}).
\end{enumerate}
For this class integrability conditions are consistently satisfied.
\item For $\sigma_{2} \neq \sigma_{1}$: \\
We first note that if $\sigma_{2} \neq \sigma_{1} = \sigma_{3}$
then from (\ref{eq:in3}) we get $E_{1}=E_{3}$. Similarly $\sigma_{1} \neq
\sigma_{2}= \sigma_{3}$ implies $E_{2}=E_{3}$. Both cases falls off the
specified Type D class. So for non-vanishing shear the eigenvalues are
distinct. Furthermore from the time propagation equation (\ref{eq:in4}) we get
\begin{equation}
0=\dot{E}_{2}-\dot{E}_{1}=(\sigma_{2}-
\sigma_{1})(3E -\case{1}/{2}\rho)
\label{eq:sp9}
\end{equation}
 from which it follows that
\begin{equation}
 E=\case{1}/{6}\rho \;.
\label{eq:sp9a}
\end{equation}
Taking the time derivative of (\ref{eq:sp9a}) gives
\begin{equation}
 E(\sigma_{1}+\sigma_{2})= 0\;,
\label{eq:sp9b}
\end{equation}
from we get $E=0$ (iff $\sigma=0$) or
\begin{equation}
 (\sigma_{1}+\sigma_{2})= 0\;.
\label{eq:sp9c}
\end{equation}

By a series of three further time derivative of (\ref{eq:sp9c}) one may show
that the eigenvalues $E_{i}$ of the Weyl tensor vanish (i.e., it is type O
rather than type D), and hence is a FRW solution.
\end{enumerate}
\end{enumerate}
This proves that for irrotational dust with a purely electric type
Weyl tensor that is degenerate the shear is also degenerate
in the same plane; furthermore, the integrability conditions are satisfied.
\subsection{Type 1}
In this case, $E_{1}\neq E_{2} \neq E_{3} \neq E_{1} $.
We deduce the following properties;
\begin{enumerate}
\item From the propagation equations (\ref{eq:in3}) it follows that the shear
eigenvalues are also distinct. For if say $\sigma_{1}=\sigma_{2}$ then from
(\ref{eq:in3})
we get $E_{1}=E_{2}$ which contradicts the requirements of this class.
\item The tetrad vectors are uniquely determined.
\item From (\ref{eq:in5}) we obtain $\Gamma^{\nu}{}_{0\mu}=0$ for $\mu\neq
\nu$ and hence the spatial tetrad vectors are Fermi propagated.
\item From the new constraint (\ref{eq:prop2}\,-\,\ref{eq:prop4}),
that is the time development of the
constraint equations (\ref{eq:con18}\,-\,\ref{eq:con20}), we note that the
spatial tetrad vectors are  hypersurface orthogonal \cite{bi:BR} i.e.,
$\Gamma^{1}{}_{23}=\Gamma^{2}{}_{31}=\Gamma^{3}{}_{12} =0$.
\end{enumerate}
For this class further analysis of constraints
(\ref{eq:prop8}\,-\,\ref{eq:prop10}) is performed below.
\subsubsection{Equivalent Constraints}
We now concentrate on the new constraint (\ref{eq:prop9}). If we write the
second term on the right hand side of constraint (\ref{eq:prop9}) as
\begin{equation}
 (\sigma_{3}-\sigma_{1})\partial_{2} E_{3}=(\sigma_{3}-\sigma_{1})
\partial_{2} (E_{3} -E_{1})+(\sigma_{3}-\sigma_{1})\partial_{2} E_{1}
\label{eq:sp10}
\end{equation}
and use (\ref{eq:con22}) we obtain the constraint
\begin{eqnarray}
\lefteqn{
(E_{3}-E_{1})\partial_{2}\sigma_{1}+(\sigma_{3}-\sigma_{1})\partial_{2} E_{1}}
\nonumber \\
&=&\case{2}/{3}\Gamma^{3}{}_{32} (\sigma_{2}-\sigma_{3})(E_{2}-E_{3})
-\case{2}/{3}\Gamma^{1}{}_{12} (\sigma_{2}-\sigma_{1})(E_{2}-E_{1})\nonumber \\
&-&\Gamma^{1}{}_{12}\left[ (\sigma_{3}-\sigma_{1}) (E_{2}-E_{1})
+(\sigma_{2}-\sigma_{1}) (E_{3}-E_{1}) \right]\;.
\label{eq:sp11}
\end{eqnarray}
Similar manipulations of the first term on the r.h.s.  of (\ref{eq:prop9})
yields the following constraint:
\begin{eqnarray}
\lefteqn{
 (E_{3}-E_{1})\partial_{2}\sigma_{3}+(\sigma_{3}-\sigma_{1})\partial_{2} E_{3}}
\nonumber \\
&=&\case{2}/{3}\Gamma^{3}{}_{32} (\sigma_{2}-\sigma_{3})(E_{2}-E_{3})
-\case{2}/{3} \Gamma^{1}{}_{12} (\sigma_{2}-\sigma_{1})(E_{2}-E_{1})\nonumber\\
&-& \Gamma^{3}{}_{32}[ (\sigma_{3}-\sigma_{1}) (E_{2}-E_{3})
+(\sigma_{2}-\sigma_{3}) (E_{3}-E_{1}) ]\;.
\label{eq:sp12}
\end{eqnarray}
Each of equations (\ref{eq:sp11},\ref{eq:sp12}) is equivalent
to the constraint (\ref{eq:prop9}), if we assume the remaining constraint
equations to be satisfied.
\subsubsection{Consistency test}
To test consistency of the two constraints (\ref{eq:sp11},\ref{eq:sp12}) we
simplify as follows: Let
\begin{equation}
X_{1} = \partial_{2}\sigma_{1}; \;\;\;
X_{3} = \partial_{2}\sigma_{3}; \;\;\;
Y_{1} = \partial_{2}E_{1}; \;\;\;
Y_{3} = \partial_{2}E_{3}\;,
\label{eq:sp13}
\end{equation}
\begin{equation}
a = (E_{3}-E_{1}); \;\;\;\;\;\;
b = (\sigma_{3} -\sigma_{1})\;.
\label{eq:sp14}
\end{equation}
So now (\ref{eq:sp11},\ref{eq:sp12}) become
\begin{eqnarray}
a X_{1} + b Y_{1} &=& R_{1}\;, \label{eq:sp15} \\
a X_{3} + b Y_{3} &=& R_{3}\;,
\label{eq:sp16}
\end{eqnarray}
with $R_{1}$ and $R_{3}$ given by the respective r.h.s.
Two further constraints are obtained from (\ref{eq:sp15},\ref{eq:sp16})
by subtracting and adding respectively i.e.,
\begin{eqnarray}
a (X_{3} - X_{1}) + b(Y_{3} - Y_{1}) &=& R_{3} - R_{1}\;, \label{eq:sp17}\\
a (X_{3} + X_{1}) + b(Y_{3} + Y_{1}) &=& R_{3} + R_{1}\;.
\label{eq:sp18}
\end{eqnarray}
Constraint (\ref{eq:sp17},\ref{eq:sp18}) are due to the time propagation
of the constraint (\ref{eq:con22}).\\

Next we modify the original constraints as follows: multiplying
(\ref{eq:con15}) by
$a = (E_{3}-E_{1})$ and (\ref{eq:con22}) by $b = (\sigma_{3} -\sigma_{1})$
gives:
\begin{eqnarray}
a X_{3} &=& a X_{1} + L_{1}\;, \label{eq:sp19} \\
b Y_{3} &=& b Y_{1} + L_{3}\;,
\label{eq:sp20}
\end{eqnarray}
where now
$L_{1} = a\times (\mbox{rhs of } (\ref{eq:con15}))$ and $L_{2} = b\times
(\mbox{rhs of } (\ref{eq:con22}))$. Taking a hint from (\ref{eq:sp17},
\ref{eq:sp18}) we use constraint (\ref{eq:sp19},\ref{eq:sp20}) to formulate
the following two further constraints as;
\begin{eqnarray}
a (X_{3} - X_{1}) + b(Y_{3} - Y_{1}) &=& L_{1} + L_{3}\;, \label{eq:sp21}\\
a (X_{3} + X_{1}) + b(Y_{3} + Y_{1}) &=& 2a X_{1} + 2b Y_{1} + L_{1} + L_{3}\;,
\label{eq:sp22}
\end{eqnarray}
which  have not been time propagated. If we substitute
(\ref{eq:sp17},\ref{eq:sp18}) into (\ref{eq:sp19}), (\ref{eq:sp20})
respectively we obtain:
\begin{equation}
 L_{1} + L_{3} = R_{3} - R_{1}
\label{eq:sp23}
\end{equation}
and
\begin{equation}
2a X_{1} + 2b Y_{1} + L_{1} + L_{3}=R_{3} + R_{1}
\label{eq:sp24}
\end{equation}
which is a further set of constraints that are true if and only if (61,62)
hold. However equation (\ref{eq:sp24}) may be
reduced to equation  (\ref{eq:sp23}) on using (\ref{eq:sp15}). This reduction
is expected since the two equations (\ref{eq:sp17},\ref{eq:sp18})
result from one constraint, namely (\ref{eq:prop9}). Thus all we did here was
to split one equation (\ref{eq:prop9}) into two equivalent new constraints
(\ref{eq:sp19},\ref{eq:sp20}). And  by inserting  original constraints
relevantly modified as (\ref{eq:sp21},\ref{eq:sp22}) into the
(\ref{eq:sp17},\ref{eq:sp18}) and simplifying we obtain one new constraint.\\

More explicitly  the new constraint (\ref{eq:sp23}) can be written as
\begin{eqnarray}
\lefteqn{
(E_{3}-E_{1})\left[\Gamma^{3}{}_{32}(\sigma_{2}- \sigma_{3})
 - \Gamma^{1}{}_{12}(\sigma_{2} -\sigma_{1})\right]} \nonumber \\
&+& (\sigma_{3} -\sigma_{1})\left[\Gamma^{3}{}_{32}(E_{2}- E_{3})
 - \Gamma^{1}{}_{12}(E_{2} -E_{1})\right] \nonumber \\
&=& \case{2}/{3}\Gamma^{3}{}_{32} (\sigma_{2}-\sigma_{3})(E_{2}-E_{3})
-\case{2}/{3}\Gamma^{1}{}_{12} (\sigma_{2}-\sigma_{1})(E_{2}-E_{1})\nonumber \\
&-& \Gamma^{1}{}_{12}\left[(\sigma_{3}-\sigma_{1}) (E_{2}-E_{1})
+(\sigma_{2}-\sigma_{1}) (E_{3}-E_{1})\right]\nonumber \\
&-& \case{2}/{3}\Gamma^{3}{}_{32} (\sigma_{2}-\sigma_{3})(E_{2}-E_{3})
+\case{2}/{3}\Gamma^{1}{}_{12} (\sigma_{2}-\sigma_{1})(E_{2}-E_{1})\nonumber \\
&+& \Gamma^{3}{}_{32}\left[ (\sigma_{3}-\sigma_{1}) (E_{2}-E_{3})
+(\sigma_{2}-\sigma_{3}) (E_{3}-E_{1}) \right]
\label{eq:sp25}
\end{eqnarray}
and reduces easily to the identity $0 =0$; which thus shows that (46),
the basis on which these equations were derived, must be identically true
(in virtue of all the other constraints that hold). A similar approach for the
constraints (\ref{eq:prop8}) and (\ref{eq:prop10}) yields
similar identities $0=0$. And hence the set of integrability conditions
(45-47) are also satisfied for Type I fields.

\section{Coordinates}
The vectors of the eigen\,-\,tetrad are hypersurface orthogonal
for both type D and type I spacetimes. So now using the tetrad\,-\'coordinate
relations (\ref{eq:in6a},\ref{eq:in6b}) there exists a coordinate system
in which line element has  diagonal form;
\begin{equation}
ds^{2} = -V^{2}dt^{2} + A^{2}dx^{2} + B^{2}dy^{2} + C^{2}dz^{2}
\label{eq:sp28}
\end{equation}
where $A,B,C,V$ are functions of the spacetime variables (Barnes and
Rowlingson, \cite{bi:BR}).
For type D spacetimes, a suitable rescaling  of the $t$ coordinate  to
set $V=1$ and use of (48) and (52) gives
\begin{equation}
ds^{2} = -dt^{2} + A^{2}(x^i)(dx^{2} + k(x,y)^{2}dy^{2}) +
\sigma^{-2}(x^j) g^{2}(z,t)dz^{2}
\label{eq:sp29}
\end{equation}
where $\sigma^2(x^i)$ is the magnitude of the shear;
this has the form assumed by Szekeres \cite{bi:SZK}.\\
\section{Conclusion}
The result that the tetrad vectors are hypersurface orthogonal appears
as an integrability condition from the $\dot{H}$ constraint with $\mu=\nu$,
and for the given tetrad choice this condition is satisfied. This has been
emphasized in \cite{bi:BR}. We have managed to show here that the remaining
$\dot{H}$ constraints for $\mu\neq \nu$ are also satisfied without imposing
any further geometric requirement on the tetrad. We have also checked that
the time derivatives of all the other constraints are identically satsified
in view of the set of constraint equations that hold under our given
conditions.\\

Thus all the constraints resulting from the vanishing of $H_{ab}$ can be
consistently satisfied in solutions with $p = 0 = \omega_{ab}$.
More specifically, $H_{ab} = 0$ is equivalent to (6); the time derivative of
this equation gives (10); if both of these equations are satisfied, there are
no further consistency conditions to be satisfied resulting from
$H\,\ddot{}_{ab} = 0$ or higher time derivatives. \\

The latter result was assumed to be valid  in the papers by Mataresse
{\it et al} \cite{bi:MPS1} and Bruni {\it et al} \cite{bi:BMP}, where on
neglecting the effect of surrounding matter, as carried by
$H_{ab}$ (in the absence of a pressure gradient), the model evolved as a
silent universe. \\

We can look at this from the viewpoint of initial data: for initial
conditions that satisfy the constraint equation $H_{ab}=0$ (that is,
(6)) at some initial time $t = t_0$, together with the constraint equation
$\dot{H}_{ab} = 0$ (that is, (10)) at that time, then for later times
(within the Cauchy development of this initial data) the magnetic part of the
Weyl tensor will remain zero. Thus vanishing of $H_{ab}$ is a consistent
condition in an open domain. It is a a dynamical restriction which will
remain valid through the evolution of the system; thus for example
the collapse of fluid elements to a curvature singularity in the form
of a Zeldovich pancake can be consistently represented by such solutions.

\subsection*{Acknowledgements}
WML would like to thank the University of Cape Town for hospitality
during the preparation of this work. We thank the FRD (South Africa)
for financial support.
\appendix
\section{Jacobi Identities}
Following [3], the commutators $[e_{a},e_{b}]$ can be written as:
\begin{equation}
[e_{a},e_{b}] = \gamma^{c}{}_{ab}e_{c}\;,
\;\;\;a,b,c=0 \ldots 3\;,
\end{equation}
where the structure constants $\gamma^{c}{}_{ab}$
are related to the Ricci coefficients $\Gamma^{c}{}_{ab}$ by:
\begin{equation}
\gamma^{c}{}_{ab} = \Gamma^{c}{}_{ab} -\Gamma^{c}{}_{ba}
\end{equation}
and inversely by:
\begin{equation}
\Gamma_{abc}=
\case{1}/{2}(\gamma_{abc}+ \gamma_{cab} -\gamma_{bca})\;.
\end{equation}
The Jacobi identity
\begin{equation}
[e_{b},[e_{c},e_{d}]] + [e_{d},[e_{b},e_{c}]] + [e_{c},[e_{d},e_{b}]]=0
\end{equation}
can then be written in terms of $\gamma^{c}_{}{ab}$ as:
\begin{equation}
\left( \begin{array}{c} f \\ bcd\end{array}  \right): \;
 ~~~~~~ \partial_{[d}\gamma^{f}{}_{cb]}
 + \gamma^{g}{}_{[dc}\gamma^{f}{}_{b]g} =0\;.
\end{equation}
That is:
\begin{equation}
\left( \begin{array}{c} 1 \\ 123\end{array}  \right): \;
\partial_{1}\gamma^{1}{}_{32}  - \partial_{2}\gamma^{1}{}_{31} +
\partial_{3}\gamma^{1}{}_{21} =
-\gamma^{2}{}_{32}\gamma^{1}{}_{21} - \gamma^{3}{}_{23}\gamma^{1}{}_{31}
+\gamma^{1}{}_{32}(\gamma^{2}{}_{12} - \gamma^{3}{}_{13})\;,
\label{eq:a1}
\end{equation}

\begin{equation}
\left( \begin{array}{c} 2 \\ 123\end{array}  \right): \;
- \partial_{1}\gamma^{2}{}_{32}  + \partial_{2}\gamma^{2}{}_{13}
- \partial_{3}\gamma^{2}{}_{12} =
- \gamma^{1}{}_{31}\gamma^{2}{}_{12} + \gamma^{3}{}_{13}\gamma^{2}{}_{32}
+ \gamma^{2}{}_{13}(\gamma^{3}{}_{23} + \gamma^{1}{}_{21})\;,
\label{eq:a2}
\end{equation}

\begin{equation}
\left( \begin{array}{c} 3 \\ 123\end{array}  \right): \;
- \partial_{1}\gamma^{3}{}_{23}  + \partial_{2}\gamma^{3}{}_{13}
- \partial_{3}\gamma^{3}{}_{12} =
\gamma^{1}{}_{21}\gamma^{3}{}_{13} - \gamma^{2}{}_{12}\gamma^{3}{}_{23}
- \gamma^{3}{}_{12}(\gamma^{1}{}_{31} + \gamma^{2}{}_{32})\;,
\label{eq:a3}
\end{equation}

\begin{equation}
\left( \begin{array}{c} \mu \\ 0 \mu \nu \end{array}  \right): \;
\partial_{0}\gamma^{\mu}{}_{\nu\mu} =
-\partial_{\nu}\theta_{\mu}
-\theta_{\nu}\gamma^{\mu}{}_{\nu\mu}\;,
\label{eq:a4}
\end{equation}

\begin{equation}
\left( \begin{array}{c} 1 \\ 023\end{array}  \right): \;
\partial_{0}\gamma^{1}{}_{32}=
\gamma^{1}{}_{32}(\theta_{1} -\theta_{2}-\theta_{3})\;,
\label{eq:a5}
\end{equation}

\begin{equation}
\left( \begin{array}{c} 2 \\ 031\end{array}  \right): \;
\partial_{0}\gamma^{2}{}_{13}=
 \gamma^{2}{}_{13}(-\theta_{1} +\theta_{2}-\theta_{3})\;,
\label{eq:a6}
\end{equation}

\begin{equation}
\left( \begin{array}{c} 3 \\ 012\end{array}  \right): \;
\partial_{0}\gamma^{3}{}_{12} =
\gamma^{3}{}_{12}(-\theta_{1} -\theta_{2}+\theta_{3})\;.
\label{eq:a7}
\end{equation}
\section{Sample Calculations}
We consider here the time propagation of the $\dot{H}$ constraint.
Starting from  the covariant form [see equation (10)]
\begin{equation}
\;\; \dot{H}: \;\;\;\;\;h^{(i}{}_{n}\eta^{j)klm}u_{k}E^{n}{}_{l;m} = 0
\label{ap:b1}
\end{equation}
and  using
\begin{equation}
(E^{n}{}_{l;m}\dot{)}=(\dot{E}^{n}{}_{l})_{;m}-E^{n}{}_{;p}u^{p}{}_{;m}
- R^{n}{}_{mpq}E^{q}{}_{l}u^{p}+R^{q}{}_{mpl}E^{n}{}_{q}u^{p}\;,
\label{ap:b3}
\end{equation}
we obtain as the time derivative of (90),
\begin{equation}
0=h^{(i}{}_{n}\eta^{j)klm}u_{k}\left[(\dot{E}^{n}{}_{l})_{;m}
-E^{n}{}_{l;p}u^{p}{}_{;m}- R^{n}{}_{mpq}E^{q}{}_{l}u^{p}+ R^{q}{}_{mpl}
E^{n}{}_{q}u^{p}
\right]\;.
\label{ap:b4}
\end{equation}

The two terms involving the Riemann tensor $R^{a}{}_{bcd}$ may be shown
to be zero as follows:
\begin{eqnarray}
&\mbox{}& h^{(i}{}_{n}\eta^{j)klm}u_{k}\left[
 R^{q}{}_{mpl}E^{n}{}_{q}u^{p} - R^{n}{}_{mpq}E^{q}{}_{l}u^{p}
  \right]  \nonumber \\
&=& \case{1}/{2} \eta^{jklm}u_{k}\left[
  R^{q}{}_{mpl}E^{i}{}_{q}u^{p} - R^{i}{}_{mpq}E^{q}{}_{l}u^{p}\right]
  \nonumber \\
&\mbox{}& + \case{1}/{2} \eta^{iklm}u_{k}\left[
  R^{q}{}_{mpl}E^{j}{}_{q}u^{p} - R^{j}{}_{mpq}E^{q}{}_{l}u^{p}
  \right] \nonumber\\
&=& \case{1}/{2}(Term\,1 + Term\,2) \;.
\label{ap:b5}
\end{eqnarray}
where
\begin{eqnarray}
Term\,1 &=& \case{1}/{2} \eta^{jklm}u_{k}\left[
  R^{q}{}_{mpl}E^{i}{}_{q}u^{p} - R^{i}{}_{mpq}E^{q}{}_{l}u^{p}\right]
  \nonumber \\
Term\,2&=& \case{1}/{2} \eta^{iklm}u_{k}\left[
  R^{q}{}_{mpl}E^{j}{}_{q}u^{p} -
  R^{j}{}_{mpq}E^{q}{}_{l}u^{p}\right]\;.
\label{ap:b6}
\end{eqnarray}
Consider now
\begin{eqnarray}
Term\,1 &=& \case{1}/{2} \eta^{jklm}u_{k}\left[
  R^{q}{}_{mpl}E^{i}{}_{q}u^{p} - R^{i}{}_{mpq}E^{q}{}_{l}u^{p}\right]
  \nonumber \\
      &=& T1 + T2
\label{ap:b7}
\end{eqnarray}
where we set
\begin{eqnarray}
T1= \eta^{jklm}E^{q}{}_{l}R^{i}_{mpq}u_{k}u^{p}\; ;
\label{ap:b8} \\
T2= \eta^{jklm}E^{i}{}_{q}R^{q}_{mpl}u_{k}u^{p}
\label{ap:b8a}\;.
\end{eqnarray}

We write the Riemann tensor $R_{smpl}$  in the
terms of the Weyl tensor as

\begin{eqnarray}
R_{smpl}&=& C_{smpl} + \case{1}/{2}(g_{sp}R_{lm} + g_{sl}R_{pm}
              -g_{mp}R_{sl} + g_{ml}R_{sp})
  \nonumber \\
     &\mbox{}& - \frac{R}{6}( g_{sp}g_{lm}-g_{sl}g_{pm})
\label{ap:b9}
\end{eqnarray}
where
\begin{equation}
C_{smpl}
\equiv (\eta_{smab}\eta_{plcd} + g_{smab}g_{plcd})u^{a}u^{c}E^{bd},
\label{ap:b10}
\end{equation}
and
\begin{equation}
g_{smab} \equiv g_{sa}g_{mb} - g_{sb}g_{ma}.
\label{ap:b11}
\end{equation}

{\bf Calculation of T1}: Using (\ref{ap:b9}) in (\ref{ap:b8}) we get
\begin{eqnarray}
T1&=& \eta^{jklm}E^{q}{}_{l}R^{i}{}_{mpq}u^{p}u_{k}
  \nonumber \\
&=& \eta^{jklm}E^{q}{}_{l}u^{p}u_{k}[C^{i}{}_{mpq}
+ \case{1}/{2}(g^{i}{}_{p}R_{qm} + g^{i}{}_{q}R_{pm}
              -g_{mp}R^{i}_{q} + g_{mq}R^{i}_{p})
  \nonumber \\
&\mbox{}& - \frac{R}{6}(g^{i}{}_{p}g_{qm} - g^{i}{}_{q}g_{pm})]
  \nonumber \\
&=& T1a + T1b
\label{ap:b12}
\end{eqnarray}
where
\newpage
\begin{eqnarray}
T1a & =& \eta^{jklm}E^{q}{}_{l}u^{p}u_{k}C^{i}{}_{mpq}
  \nonumber \\
&=& E^{q}{}_{l}E^{bd}u^{a}u^{c}u^{p}u_{k}[
\eta^{jklm}\eta^{i}{}_{mab}\eta_{pqcd}
+\eta^{jklm}(g^{i}{}_{a}g_{mb}-g^{i}{}_{b}g_{ma})g_{pqcd}]
  \nonumber \\
&=& E^{q}{}_{l}E^{bd}
\eta^{jklm}\eta^{i}{}_{mab}\eta_{pqcd}u^{a}u^{c}u^{p}u_{k}
+ E^{q}{}_{l}E^{d}{}_{m} \eta^{jklm} u^{i}u^{c}u^{p}u_{k} g_{pqcd}
  \nonumber \\
&\mbox{}& +E^{q}{}_{l}E^{id} \eta^{jklm} u_{m}u^{c}u^{p}u_{k}g_{pqcd}
  \nonumber \\
&=& E^{q}{}_{l}E^{d}{}_{m} \eta^{jklm} u^{i}u^{c}u^{p}u_{k}
(g_{pc}g_{qd} - g_{pd}g_{qc})
 \nonumber \\
&=&  E_{ld}E^{d}{}_{m}\eta^{jklm}u^{i}u_{k}u_{p}u^{p}
  - E_{lc}E_{mp}\eta^{jklm}u^{i}u^{c}u^{p}u_{k}
  \nonumber \\
&=&0 \;\;\;  \mbox{ (Symmetry properties)}
\label{ap:b13}
\end{eqnarray}
and
\begin{eqnarray}
T1b &=& \eta^{jklm}E^{q}{}_{l}u^{p}u_{k}[
 \case{1}/{2}(g^{i}{}_{p}R_{qm} + g^{i}{}_{q}R_{pm}
              -g_{mp}R^{i}_{q} + g_{mq}R^{i}_{p})
  \nonumber \\
&\mbox{}&- \frac{R}{6}(g^{i}{}_{p}g_{qm} - g^{i}{}_{q}g_{pm})]
\nonumber \\
&=&\eta^{jklm}u_{k}[
\case{1}/{2}(E^{q}{}_{l}R_{qm}u^{i} + E^{i}{}_{l}R_{pm}u^{p})
-\case{1}/{2}(E^{q}{}_{l}R^{i}{}_{q}u_{m} - E_{ml}R^{i}{}_{p}u^{p})
\nonumber \\
&\mbox{}& - \frac{R}{6}(E_{ml}u^{i} - E^{i}{}_{l} u_{m}) ].
\label{ap:b14}
\end{eqnarray}
For dust
\begin{equation}
R_{pm} = \case{1}/{2}\rho(h_{pm}+u_{p}u_{m}).
\label{ap:b15}
\end{equation}
and hence  (\ref{ap:b14}) becomes
\begin{eqnarray}
T1b &=& -\case{1}/{2}\rho E^{i}{}_{l}\eta^{jklm} u_{m}u_{k}
    -\frac{R}{6}(E_{ml}u^{i}-E^{i}{}_{l}u_{m})\eta^{jklm}u_{k}
  \nonumber \\
& = & 0 \;\;\;  \mbox{ (Symmetry properties)}\;.
\label{ap:b16}
\end{eqnarray}
So now
$T1= T1a+T1b = 0$.\\

{\bf Calculation of T2}: Using (\ref{ap:b9}) in (\ref{ap:b8a}) we get
\begin{eqnarray}
T2&=& \eta^{jklm}E^{i}{}_{q}R^{q}{}_{mpl}u_{k}u^{p}
  \nonumber \\
&=& \eta^{jklm}E^{i}{}_{q}u_{k}u^{p}[
C^{q}{}_{mpl}
+ \case{1}/{2}(g^{q}{}_{p}R_{lm} + g^{q}{}_{l}R_{pm})
-\case{1}/{2} (g_{mp}R^{q}{}_{l} - g_{ml}R^{q}{}_{p})
\nonumber \\
&\mbox{}& -\case{1}/{6} (g^{q}{}_{p}g_{lm} -  g^{q}{}_{l}g_{pm})]
\nonumber \\
&=& T2a + T2b
\label{ap:b18}
\end{eqnarray}
and as in (\ref{ap:b12}) we set
\begin{eqnarray}
T2a &=& \eta^{jklm}E^{i}{}_{q}u_{k}u^{p} C^{q}{}_{mpl}
\nonumber \\
&=& \eta^{jklm}E^{i}{}_{q}E^{bd}u^{a}u^{c}u_{k}u^{p}[
  \eta^{q}{}_{mab}\eta_{plcd}
+ (g^{q}{}_{a}g_{mb} -g^{q}{}_{b}g_{ma}) g_{plcd}]
\nonumber \\
&=& \eta^{jklm} \eta^{q}{}_{mab}\eta_{plcd}
E^{i}{}_{q}E^{bd}u^{a}u^{c}u_{k}u^{p}
-\eta^{jklm}E^{i}{}_{b}E^{bd}u^{c}u^{p}u_{m}u_{k}
(g_{pc}g_{ld} - g_{pd}g_{lc})
\nonumber \\
&=& 0 \;\;\;  \mbox{ (Symmetry properties.)}
\label{ap:b19}
\end{eqnarray}
and
\begin{eqnarray}
T2b &=& \eta^{jklm}E^{i}{}_{q}u_{k}u^{p}[
\case{1}/{2}(g^{q}{}_{p}R_{lm} + g^{q}{}_{l}R_{pm})
-\case{1}/{2} (g_{mp}R^{q}{}_{l} - g_{ml}R^{q}{}_{p})
\nonumber \\
&\mbox{}& -\frac{R}{6} (g^{q}{}_{p}g_{lm} -  g^{q}{}_{l}g_{pm})]
\nonumber \\
&=& 0 \;\;\;  \mbox{ (Symmetry properties)}\;.
\label{ap:b20}
\end{eqnarray}
Here also $T2= T2a +T2b =0$ and hence from (\ref{ap:b7})
\begin{equation}
Term\,1 =0 \;.
\label{ap:b21}
\end{equation}

A similar program of calculations with $i$ and $j$ interchanged gives
$Term\,2 = 0$ and the equation (\ref{ap:b4}) takes the final form
\begin{equation}
0=h^{(i}{}_{n}\eta^{j)klm}u_{k}\left[(\dot{E}^{n}{}_{l})_{;m}
-E^{n}{}_{l\; ;p}u^{p}{}_{;m}
\right]\;.
\label{ap:b21a}
\end{equation}

In tetrad form (\ref{ap:b21a}) becomes
\begin{eqnarray}
0&= &h^{(\mu}{}_{\tau} \eta^{\nu) \kappa\alpha\beta}
u_{\kappa}\left[\partial_{\beta}\dot{E}^{\tau}{}_{\alpha}
-\theta^{p}_{\beta}\partial_{p} E^{\tau}{}_{\alpha;p}+
\Gamma^{\tau}{}_{\beta\epsilon}\dot{E}^{\epsilon}{}_{\alpha}
-\Gamma^{\epsilon}{}_{\beta\alpha} \dot{E}^{\tau}{}_{\epsilon}\right.
\nonumber \\
&-&\left.\theta^{p}_{\beta}(\Gamma^{\tau}{}_{p\epsilon}E^{\epsilon}{}_{\alpha}
-\Gamma^{\epsilon}{}_{p\alpha}E^{\tau}{}_{\epsilon})\right]\;.
\label{ap:b22}
\end{eqnarray}
Setting $\mu=\nu$ in (\ref{ap:b22}) yields:
\begin{equation}
0= \eta^{\mu\kappa\alpha\beta} u_{\kappa}\Gamma^{\mu}{}_{\beta\alpha}
\left[(\dot{E}_{\alpha}-\dot{E}_{\mu})
-\theta_{\beta}(E_{\alpha}- E_{\mu}) \right].
\label{ap:b23}
\end{equation}

If we substitute the propagation equation (\ref{eq:in4}) in (112)
and consider the values of $\mu=1,2,3$ we obtain the equations
\begin{eqnarray}
0&=&\Gamma^{1}{}_{23}(E_{3}- E_{1})(\sigma_{2}- \sigma_{3})\;, \\
0&=&\Gamma^{2}{}_{31}(E_{1}- E_{2})(\sigma_{1}-\sigma_{3})\;,  \\
0&=&\Gamma^{3}{}_{12}(E_{2}- E_{3})(\sigma_{1}-\sigma_{2})\;.
\end{eqnarray}
which are the constraints (39-41).\\

For values $\mu \neq \nu$, say $\mu=1, \nu=3$,  (\ref{ap:b22}) becomes
\begin{eqnarray}
0&=&\partial_{2}(\dot{E}_{3}-\dot{E}_{1})-\theta_{2}\partial_{2}(E_{3}-E_{1})
+\Gamma^{1}{}_{12}\left[ (\dot{E}_{2}- \dot{E}_{1})
-\theta_{1}(E_{2}-E_{1}) \right]\nonumber \\
&-&\Gamma^{3}{}_{32}\left[(\dot{E}_{2}-\dot{E}_{3})
-\theta_{3}(E_{2}-E_{3})\right]
\label{ap:b24}
\end{eqnarray}

If we substitute the propagation equation (\ref{eq:in4}) into
(\ref{ap:b24}) and use the
original constraint equations repeatedly we obtain
\begin{eqnarray}
0&=&(E_{3}-E_{1})\partial_{2}\sigma_{1}+(\sigma_{3}-\sigma_{1})
\partial_{2}E_{3} +\case{1}/{3}\Gamma^{1}{}_{12}
(\sigma_{2}-\sigma_{1})(5E_{2}+4E_{1})\nonumber \\
&-&\case{1}/{3}\Gamma^{3}{}_{32} (E_{2}-E_{3}) (5\sigma_{2}+4\sigma_{3});
\label{ap:b25}
\end{eqnarray}
The remaining set of $\mu=2, \nu=3$ and $\mu=1, \nu=2$ yield
\begin{eqnarray}
0&=&(E_{2}-E_{3})\partial_{1}\sigma_{3}+(\sigma_{2}-\sigma_{3})
\partial_{1} E_{2}+\case{1}/{3} \Gamma^{3}{}_{31}
(\sigma_{1}-\sigma_{3}) (5E_{1}+4E_{3})\nonumber\\
&-&\case{1}/{3} \Gamma^{2}{}_{21} (E_{1}-E_{2}) (5\sigma_{1}+4\sigma_{2});
\label{ap:b26}
\end{eqnarray}
and
\begin{eqnarray}
0&=&(E_{1}-E_{2})\partial_{3}\sigma_{2}+(\sigma_{1}-\sigma_{2})
\partial_{3} E_{1}+\case{1}/{3} \Gamma^{2}{}_{23}
(\sigma_{3}-\sigma_{2}) (5E_{3}+4E_{2})\nonumber\\
&-&\case{1}/{3} \Gamma^{1}{}_{13} (E_{3}-E_{1}) (5\sigma_{3}+4\sigma_{1})
\label{ap:b27}
\end{eqnarray}
respectively, which are the constraints (45-47).



\end{document}